\documentclass{vgtc}                          % final (conference style)
\ifpdf%                                % if we use pdflatex
  \pdfoutput=1\relax                   % create PDFs from pdfLaTeX
  \usepackage{graphicx}                % allow us to embed graphics files
  \DeclareGraphicsExtensions{.pdf,.png,.jpg,.jpeg} % for pdflatex we expect .pdf, .png, or .jpg files
\else%                                 % else we use pure latex
  \usepackage{graphicx}                % allow us to embed graphics files
  \DeclareGraphicsExtensions{.eps}     % for pure latex we expect eps files
\fi%

\graphicspath{{figures/}{pictures/}{images/}{./}} % where to search for the images

\usepackage{microtype}                 % use micro-typography (slightly more compact, better to read)
\PassOptionsToPackage{warn}{textcomp}  % to address font issues with \textrightarrow
\usepackage{textcomp}                  % use better special symbols
\usepackage{mathptmx}                  % use matching math font
\usepackage{times}                     % we use Times as the main font
\usepackage{cite}                      % needed to automatically sort the references
\usepackage{tabu}                      % only used for the table example
\usepackage{booktabs}                  % only used for the table example
\usepackage[perpage]{footmisc}
\usepackage{xcolor}

\onlineid{1012}

\vgtccategory{Research}

\vgtcinsertpkg

\title{Topological Data Analysis with the ``Topology ToolKit''}
\title{Topology for Dummies with the ``Topology ToolKit''}
\title{Topological Data Analysis Made Easy with the 
Topology ToolKit}

\newcommand{\mySpace}{\vspace{-1ex}}
\renewcommand{\mySpace}{}

\newcommand{\extraSpace}{\vspace{0.25ex}}
\newcommand{\extraDoubleSpace}{\extraSpace\extraSpace}

\author{
Guillaume Favelier
\\ %
\scriptsize Sorbonne Universite %
\and 
Charles Gueunet
\\ %
\scriptsize Kitware, Sorbonne Universite %
\and
Attila Gyulassy
\\ %
\scriptsize SCI Institute, University of Utah %
\and
Julien Jomier
\\ %
\scriptsize Kitware%
\and
Joshua A. Levine
\\ %
\scriptsize University of Arizona%
\and
Jonas Lukasczyk
\\ %
\scriptsize TU Kaiserslautern%
\and
Daisuke Sakurai
\\ %
\scriptsize Zuse Institute Berlin%
\and
Maxime Soler
\\ %
\scriptsize Total, Sorbonne Universite %
\and 
Julien Tierny
\\ %
\scriptsize CNRS, Sorbonne Universite %
\and
Will Usher
\\ %
\scriptsize SCI Institute, University of Utah%
\and
Qi Wu
\\ %
\scriptsize SCI Institute, University of Utah \& UC Davis
}

\teaser{
    \centering
    \includegraphics[width=\linewidth]{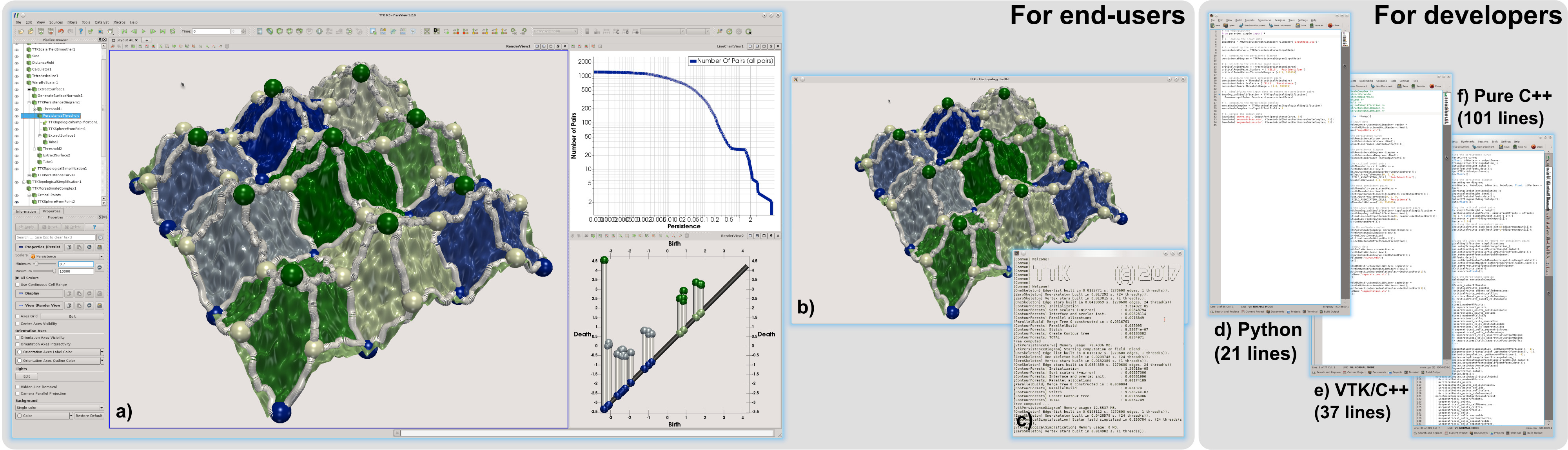}
    \caption{
    TTK is a software platform for topological data analysis in 
scientific visualization. It is both easily accessible to end users
(ParaView plugins (a), VTK-based generic GUIs (b) or command-line 
programs (c)) and flexible for developers (Python (d),
VTK/C++ (e) or dependence-free C++ (f) bindings). TTK provides an efficient and 
unified approach to topological data representation
and simplification, which enables in this example a discrete Morse-Smale 
complex 
(a) to comply to the level of simplification dictated by
a piecewise linear persistence diagram (bottom-right linked view, a). 
Code 
snippets are provided (d-f) to reproduce this pipeline.
    }
}

\begin{document}


\firstsection{Level of the tutorial}
\maketitle
This tutorial is targeted at a \emph{Beginner} audience.

\mySpace
\section{Abstract}
This tutorial presents topological methods for the analysis and 
visualization of scientific data from a user's perspective, with the 
Topology ToolKit
% \josh{``Topology ToolKit''}{quotes?} 
(TTK), a recently released open-source library for 
topological data analysis.
Topological methods %for data analysis and visualization 
have 
% \josh{considerably gained}{gained considerably} 
gained considerably
in popularity and maturity over 
the last twenty years 
and success stories of established methods have been documented in a wide range 
of applications (combustion, chemistry, astrophysics, material sciences, etc.) 
with 
% \josh{}{both} 
both
acquired 
% \josh{or}{and}
and
simulated data, in 
% \josh{a}{both} 
both
post-hoc 
% \josh{or}{and} 
and
in-situ contexts.
While reference textbooks have been published on the topic, no tutorial 
% \josh{}{at IEEE VIS} 
at IEEE VIS 
has 
covered this area in recent years, and never 
at 
% \josh{a software level}{\emph{a software level}},
\emph{a software level}
% \josh{}{and} 
and
from 
% \josh{a user's point of view}{\emph{a user's point-of-view}}.
\emph{a user's point-of-view}.
This tutorial fills this gap by providing a beginner's introduction to 
topological methods for practitioners, researchers, students, and 
lecturers. In particular, instead of focusing on theoretical aspects and 
algorithmic details, this tutorial focuses on how topological methods can be 
useful in practice for concrete data analysis tasks 
% \josh{involving}{such as}
such as
segmentation, 
feature extraction or tracking. 
% \josh{In particular, t}{T}
The tutorial describes in 
detail
% \josh{s}{} 
how to achieve these tasks with TTK. First, after an introduction to 
topological methods and their application in data analysis, a brief overview of 
TTK's main entry point for end users, namely ParaView, will be presented. 
Second, an overview of TTK's main features will be given. A running example 
will be described in 
% \josh{more details}{detail},
detail,
showcasing 
% \josh{the access to}{how to access} 
how to access
TTK's features via 
ParaView, Python, VTK/C++,
% \josh{ or}{, and} 
and
C++. Third, hands-on sessions will 
concretely show how to use TTK in ParaView for multiple, representative data 
analysis tasks. Fourth, the usage of TTK will be presented for developers, in 
particular by describing several examples of visualization and data analysis 
projects 
% \josh{which}{that}
that
were built on top of TTK. Finally, some feedback regarding 
the 
usage of TTK as a teaching platform for 
% \josh{computational topology classes}{topological analysis}
topological analysis
will be 
given. Presenters of this tutorial 
% come from various background (academic, 
% labs, industry)
include experts in topological methods, 
core authors of TTK as well as active users, coming from academia, 
labs, or
% \josh{}{,} or 
industry. A large part of the tutorial will be dedicated to hands-on 
exercises and a rich material package (including TTK pre-installs in 
virtual machines, code, data, demos, video tutorials, etc.)
% \josh{}{.}) 
will be provided to 
the participants. This tutorial mostly targets students, 
% \josh{practictioners}{practitioners} 
practitioners
and 
researchers who are not experts in topological methods but who are
interested in using them in their daily tasks. 
% \josh{It also targets}{We also target} 
We also target
researchers 
already familiar to topological methods and who are interested in using or 
contributing to TTK.

\mySpace
\section{Tutorial organization}
%   \mySubSection{Motivations}

\textbf{Motivations}
  As scientific datasets become more intricate and larger in size, advanced 
data analysis algorithms are needed for their efficient visualization and 
exploration. For scalar field visualization, topological analysis techniques 
\cite{pascucci_topoInVis10, heine16, tierny_book18}
have shown to be practical solutions in various contexts by enabling the 
concise and complete capture of the structure of the input data into high-level 
\emph{topological abstractions} such as contour trees \cite{carr00, 
gueunet_ldav16, gueunet_ldav17}, 
Reeb graphs \cite{pascucci07, biasotti08, tierny_vis09, parsa12}, or 
Morse-Smale 
complexes \cite{gyulassy_vis08, gyulassy_vis14, Defl15}.
Such topological abstractions are fundamental data structures that 
enable 
% the development of
advanced data 
analysis, exploration and visualization techniques, including for instance: 
small seed set extraction for fast isosurface traversal \cite{vanKreveld97, 
carr04, santos_escience09}, feature tracking \cite{sohn06}, transfer function 
design for volume 
rendering \cite{weber07}, 
data simplification \cite{tierny_vis12} and compression \cite{soler_pv18},
similarity estimation \cite{tierny_cgf09, thomas14}, geometry processing 
\cite{tierny_tvcg11, vintescu17},
% \josh{}{,} 
or 
application-driven segmentation and analysis tasks.
% \cite{laney_vis06, 
% gyulassy07, gyulassy_ev14, gyulassy_vis15}. 
Successful applications in a 
variety of fields of science,
% to fields of science other than computer science, 
including combustion 
\cite{laney_vis06, bremer_tvcg11, gyulassy_ev14}, 
fluid dynamics \cite{kasten_tvcg11, fangChen13},
material sciences \cite{gyulassy07, gyulassy_vis15, favelier16, Lukasczyk17},
chemistry \cite{chemistry_vis14, harshChemistry}, 
% \josh{or}{and} 
and
astrophysics \cite{sousbie11, shivashankar2016felix}, have 
% \josh{even}{} 
been documented, which further 
% \josh{stresses}{demonstrates} 
demonstrates
the importance of 
these
% class of 
techniques.

While reference textbooks have been published 
% \josh{on the topic}{}
\cite{edelsbrunner09} 
% \josh{to}{that} 
that
present the fundamental aspects of these techniques, 
no tutorial has covered this area in recent years at IEEE VIS. The
% \josh{, the}{.  The} 
latest 
tutorial related to topology 
% \josh{being}{occurred} 
occurred
nearly 10 years 
% \josh{old}{ago} 
ago
\cite{weber09}.
Moreover, despite their popularity and success in applications, topological 
methods have not yet been widely adopted as a standard data analysis tool for 
end users and developers. We believe one of the reasons for this is the lack of 
open-source software packages 
% \josh{implementing}{that implement these algorithms} 
that implement these algorithms
in a generic, 
user-friendly,
% \josh{}{,} 
and 
efficient way.
% \josh{ these algorithms}{}. 
Recently, the 
Topology 
ToolKit
% \josh{``Topology 
% ToolKit''}{quotes} 
(TTK) 
\cite{ttk17}, an open-source library for topological data analysis has been 
released (BSD license) to fill this gap. TTK is mostly written in C++ ($\sim$ 
210k lines of code) and 8 institutions have contributed to its development so 
far, including 5 academic institutions (CNRS, Sorbonne Universite, University of 
Utah, Zuse Institute, University of Arizona) and 3 private companies (Kitware, 
Total, Caboma Inc.). TTK is currently supported under Linux, 
MacOS,
% \josh{}{,}
and Windows.
Since its initial release on Github a year ago, TTK's 
website collected more than 64k page-views, from more than 6.7k unique 
visitors. These statistics indicate that a user base for TTK 
% \josh{indeed}{} 
exists and 
that further efforts towards the explanation of TTK's usage 
would be beneficial to the community.

The main motivation for this tutorial is therefore to introduce to beginners 
how topological methods can be useful 
% \josh{to analyze}{for analyzing}
for analyzing
their data, and how to do it 
with TTK.

%   \mySpace
%   \mySubSection{Target audience}

\extraDoubleSpace
\noindent
\textbf{Target audience}
  This tutorial mostly targets beginners, students, practitioners,
%   \josh{}{,} 
and 
researchers who are not experts in topological methods but who are interested in 
using them in their daily tasks. It also targets researchers already familiar 
to topological methods and who are interested in using or contributing to TTK.
  
%   \mySpace
%   \mySubSection{Tutorial goals}
\extraDoubleSpace
  \noindent
\textbf{Tutorial goals}
  The goals of this tutorial are to present the key tools in topological data 
analysis (the 
% \josh{P}{p}
Persistence diagram, the Reeb graph and its variants, the 
Morse-Smale complex, etc.)
% \josh{}{.}) 
and how they can be used in practice for 
precise data 
analysis tasks, including data segmentation and feature extraction. 
All examples will be illustrated with TTK. 
This 
tutorial also aims at presenting TTK and
its different usage modalities
% the different 
% modalities of its usage 
(ParaView, Python, VTK/C++, C++). We expect 
participants to become capable of using TTK independently, 
at least with ParaView (possibly with Python), after attending the tutorial.
  
%   \mySpace
%   \mySubSection{Hands-on material}
\extraDoubleSpace
\noindent
\textbf{Hans-on material}
  A large part of the tutorial will be dedicated to hands-on exercises with TTK 
and ParaView \cite{paraviewBook}. We will provide a rich material package 
including TTK pre-installs in virtual machines (to be used by attendees during 
the tutorial), code, data, demos, video tutorials, etc. Most of this material is 
already available on TTK's website \cite{ttk17}.  
% \josh{}{
Our idea is that
participants who bring a laptop  will be able to follow along,
regardless of their native OS.  Attendees who attend just to listen and learn
will also benefit from the tutorial and receive sufficient material to try out
our examples at home.
% } 

%   \mySpace
%   \mySubSection{Qualification of the presenters}
\extraDoubleSpace
\noindent
\textbf{Qualification of the presenters}
  Presenters include experts in topological methods, who 
%   \josh{}{have}
  have
published many papers on the topic in premier venues as well as textbooks 
(\autoref{sec_background}). 
The list of presenters also includes the 
core authors of TTK and its most active users.
  
%   \mySpace
%   \mySubSection{Proposal strengths}
\extraDoubleSpace
\noindent
\textbf{Proposal strengths}
  In contrast to previous tutorials on topological methods \cite{weber09}, 
we 
believe this proposal to have a unique 
% \josh{concerete}{concrete}
concrete
and applicative appeal, by its 
focus on the \emph{usage} of topological methods rather than on their 
\emph{foundations}. Thus, we expect it to attract a larger audience than the 
specific subset of IEEE VIS attendees typically found in traditional topology 
sessions.
% \josh{topology crowd}{
% subset of the IEEE VIS audience who attends topology sessions.
% on topology
% analysis.
% }.
Also, this tutorial is unique in the sense that TTK has never 
been presented before in a tutorial.
 
We believe that the list of presenters is also a strength of this 
proposal. First, it includes topology experts as well as core developers and 
users of TTK. More importantly, it includes researchers with a variety of 
experience profiles (Ph.D. students, post-docs, professors) and backgrounds 
(industry, labs, academia), which will ease interactions with a potentially 
heterogeneous audience. Moreover, the particularly large number of presenters 
(11) has two merits. First, it imposes a mini-symposium
% \josh{simposium}{symposium} 
structure, where 
speakers will give presentation lasting between 10 and 20 minutes, which will 
result in a lively 
% \josh{rythm}{rhythm}
rhythm
in the overall tutorial. Second, this large number of 
presenters will be instrumental during the hands-on exercises, as there will be 
enough presenters such that one presenter can assist a small group of attendees 
(typically 3 to 4).

Finally, we believe the detailed program of the tutorial (see 
\autoref{sec_details}) achieves a 
% \josh{fine}{} 
balance between concepts, usage 
descriptions and application examples.
  
%   \mySpace
%   \mySubSection{Detailed content}
\extraDoubleSpace
\noindent
\textbf{Detailed content}
  \label{sec_details}
  The tutorial is divided into three main parts (each part being subdivided 
into modules), for a target duration of 
approximately 3 hours. These three groups of modules can be organized 
differently to
% The tentative content is as follows.  
% \josh{}{
% While the following organization 
% follows 
% is a logical partitioning, we believe these modules can be partitioned 
% differently if needed,
to fit any standard structure for breaks to match the tutorial schedule of IEEE
VIS.
% }

\extraSpace

\noindent
\textbf{A. Preliminaries (60 minutes)}

\noindent
\underline{A1. General introduction} \emph{(5 minutes, by Julien Tierny)}

\noindent
\underline{A2. Introduction to topological methods for data analysis} \emph{(30 
minutes, by Attila Gyulassy)} This talk will present the core tools in 
topological data analysis (the 
% \josh{P}{p}
Persistence diagram \cite{edelsbrunner09}, the 
Reeb graph and its variants \cite{carr00, gueunet_ldav17, biasotti08, 
pascucci07, tierny_vis09, parsa12}, the Morse-Smale complex 
\cite{gyulassy_vis08, gyulassy_vis14, Defl15}). In particular, it will detail 
how these tools can be used for data segmentation and feature extraction.

\noindent
\underline{A3. Quick introduction to ParaView's user interface}
\emph{(25 minutes, 
by Julien Jomier)} This talk will provide a brief description of 
ParaView's
% \josh{}{'s} 
main 
interface \cite{paraviewBook}, in order to support its usage for beginners in 
the subsequent hands-on session. This will cover the usage of filters, pipeline 
design and view manipulation, state files backups and Python exports.
% pipeline composition and view 
% manipulation as well as the usage of filters.

\extraSpace

\noindent
\textbf{B. Hands-on 
% \josh{exercices}{exercises} 
exercises
(70 minutes)}

\noindent
\underline{B1. General usage of TTK} \emph{(10 minutes, by Julien Tierny)}
This talk will briefly describe TTK's usage philosophy. It will 
 briefly present how TTK can be used from ParaView, Python, VTK/C++ or C++.

  \noindent
  \underline{B2. Segmenting medical data with merge trees} \emph{(20 minutes, 
by 
Charles Gueunet)} This hands-on TTK/ParaView exercise will be a step-by-step 
tutorial showing how to extract individual bones in a medical CT scan 
interactively with merge trees.

  \noindent
  \underline{B3. Extracting filament structures with the Morse-Smale complex} 
\emph{(20 minutes, by Guillaume Favelier)} This hands-on TTK/ParaView exercise 
will show step-by-step  how to extract filament structures 
with the Morse-Smale complex on chemistry data.

  \noindent
  \underline{B4. Topology-aware data compression} \emph{(20 minutes, by Maxime 
Soler)} This hands-on TTK/ParaView exercise 
will show step-by-step  how to compress data while guaranteeing feature 
preservation.

\extraSpace

\noindent
\textbf{C. Advanced usage (60 minutes)}

  \noindent
  \underline{C1. TTK's architecture and core data structures} \emph{(10 minutes, 
by Will Usher)} This talk will present TTK's architecture.

  \noindent
  \underline{C2. Topology driven volume rendering with TTK} \emph{(10 minutes, 
by Qi Wu)} This talk will present topo-vol \cite{topo-vol}, an implementation 
of topology-driven volume rendering \cite{weber07} built on top of TTK.
  
  \noindent
  \underline{C3. Advanced data analysis with TTK} \emph{(10 minutes, by Jonas 
Lukasczyk)} This talk will describe the processing of the SciVis 2018 contest 
data with TTK.
  
  \noindent
  \underline{C4. Bivariate data analysis with TTK} \emph{(10 minutes, by 
Daisuke Sakurai)} This talk will describe a user interface for bivariate data 
exploration built on top of TTK.
  
  \noindent
  \underline{C5. TTK as a teaching platform} \emph{(15 minutes, by Joshua 
Levine)} This talk will provide feedback about our experience in using TTK in 
our topological data analysis classes.
  
  \noindent
  \underline{C6. Concluding remarks} \emph{(5 minutes, by Julien Tierny)}
  
\mySpace
\section{Background and contact information}
\label{sec_background}

\textbf{Guillaume Favelier} 
\href{mailto:guillaume.favelier@sorbonne-universite.fr}{
-- \emph{guillaume.favelier@sorbonne-universite.fr} --}
is a research engineer at Sorbonne Universite since late 2015.
He received the master degree in 
Image Processing
% Computer Science
from Sorbonne Universite, also in 2015.
His notable contributions to TTK include the implementation of the Morse-Smale 
complex and the implicit triangulation.
Guillaume regularly demonstrates TTK capabilities through live demos and 
tutorials.
% occasionally shows
% the capabilities of TTK through demonstrations and tutorials.
% Guillaume's
His research interests include 
discrete geometry, rendering algorithms and visualization methods.

\extraSpace

\noindent
\textbf{Charles Gueunet}
\href{mailto:charles.gueunet@kitware.com}{
-- \emph{charles.gueunet@kitware.com} --}
is a Ph.D. student with Kitware and Sorbonne Universite. 
% Prior to joining 
% Kitware, 
Charles received the engineering degree in 2015 from EISTI.
His notable contributions to TTK include the implementation of the merge tree 
features.

\extraSpace

\noindent
\textbf{Attila Gyulassy}
\href{mailto:jediati@sci.utah.edu}{
-- \emph{jediati@sci.utah.edu} --} 
% todo.
received the bachelor’s of Arts in computer science and applied mathematics
from the University of California, Berkeley in 2003 and the PhD degree in 
computer science from the University of California, Davis in 2009.
His research interests as a research scientist at the Scientific Computing and 
Imaging (SCI) Institute, University of Utah, include topology-based data 
analysis and visualization.

\extraSpace

\noindent
\textbf{Julien Jomier}
\href{mailto:julien.jomier@kitware.com}{
-- \emph{julien.jomier@kitware.com} --} 
is currently
directing Kitware's European subsidiary in 
Lyon, France. Julien received both his B.S. and M.S in Electrical 
Engineering and Information Processing in 2002 from the ESCPE-Lyon 
(France) and an M.S. in Computer Science from The University of North 
Carolina at Chapel Hill (UNC) in 2003. He worked on a variety of 
projects in the areas of parallel and distributed computing, mobile 
computing, image processing, and visualization. Prior to joining 
Kitware, Mr. Jomier was a Faculty Research Lecturer of Radiology at 
UNC
% the University of North Carolina 
and a member of the Computer-Aided 
Diagnosis and Display Laboratory.

\extraSpace

\noindent
\textbf{Joshua A. Levine}
\href{mailto:josh@email.arizona.edu}{
-- \emph{josh@email.arizona.edu} --} 
is an assistant professor in the Department of Computer Science at University 
of Arizona. Prior to starting at Arizona, he was an assistant professor at 
Clemson University, and before that a postdoctoral research associate at the 
University of Utah's SCI Institute.  He received his PhD from The Ohio State 
University.
% \josh{ after completing his BS and MS in Computer Science from Case 
% Western Reserve University}{}. 
% \josh{
His research interests include 
visualization, geometric modeling, topological analysis, mesh generation, vector 
fields, performance analysis, and computer graphics.
% }{}

\extraSpace

\noindent
\textbf{Jonas Lukasczyk}
\href{mailto:jl@jluk.de}{
-- \emph{jl@jluk.de} --} 
% todo.
is a PhD student at the Technical University of Kaiserslautern.
His research interests include interactive 
visual analytics with topological methods.

\extraSpace

\noindent
\textbf{Daisuke Sakurai}
\href{mailto:sakurai@zib.be}{
-- \emph{sakurai@zib.be} --} 
is a postdoctoral researcher at Zuse Institute Berlin
(ZIB). He received his Ph.D.~at the University of Tokyo in 2015, at which
time he was also a special student at the Japan Atomic Energy Agency
(JAEA). After graduation he worked at JAEA, Institute of Physical and
Chemical Research (aka RIKEN), and the Computer Science Department (LIP6)
of Sorbonne Universite. From April 2017 he is at the current position
working for the project \emph{High Definition Clouds and Precipitation for
advancing Climate Prediction} (HD(CP)$^2$). His research interests include
multi-field visualization, atmospheric sciences, topological analysis, and
mathematical visualization.

\extraSpace

\noindent
\textbf{Maxime Soler}
\href{mailto:soler.maxime@total.com}{
-- \emph{soler.maxime@total.com} --} is a Ph.D. student with Total and 
Sorbonne Universite.
Maxime received the engineering degree in 2015 from EISTI.
His notable contributions to TTK include the implementation of topology-aware
compression and distances between persistence diagrams.

\extraSpace

\noindent
\textbf{Julien Tierny}
\href{mailto:julien.tierny@sorbonne-universite.fr}{
-- \emph{julien.tierny@sorbonne-universite.fr} --} 
received the Ph.D. degree in Computer Science from Lille 1 
University in 2008 and the Habilitation degree (HDR) from Sorbonne Universite
in 2016. He is currently a CNRS permanent research scientist, affiliated 
with Sorbonne Universite since September 2014 and with Telecom ParisTech from 
2010 to 2014. Prior to his CNRS tenure, he held a Fulbright fellowship (U.S. 
Department of State) and was a post-doctoral research associate at the 
Scientific Computing and Imaging Institute at the University of Utah. His 
research expertise includes topological data analysis for scientific 
visualization. 
% Dr. Julien Tierny received several awards for his research, 
% including best paper awards. 
He is the lead developer of the Topology ToolKit 
(TTK).
% , an open source library for topological data analysis.

\extraSpace

\noindent
\textbf{Will Usher}
\href{mailto:will@sci.utah.edu}{
-- \emph{will@sci.utah.edu} --} 
is a PhD student at the SCI Institute at the University of Utah.
His research interests include interactive ray tracing, virtual reality,
and distributed rendering. He has used TTK to develop a topology guided
volume visualization tool, and recently rewritten TTK's build system to
follow a modern CMake approach, making it easier to use.
 
\extraSpace

\noindent
\textbf{Qi Wu}
\href{mailto:qadwu@ucdavis.utah.edu}{
-- \emph{qadwu@ucdavis.edu} --} 
% is a PhD student at UC Davis.
% His research interests include interactive ray tracing, virtual reality,
% and distributed rendering. He has used TTK to develop a topology guided
% volume visualization tool.
% Qi WU 
is a PhD candidate at the Computer Science Department of the University of 
California, Davis. Prior to UC Davis, he obtained his Master degree from the SCI 
institute, University of Utah. His research interests include scientific 
visualization, ray tracing and distributed rendering on large scale systems. 
Together with Will Usher, he has used TTK to develop TopoVol, which is a 
topology guided volume exploration tool enabling sophisticated volume 
classification and rendering.

% \vspace{-2ex}

% \firstsection{Motivation}
% \maketitle

% The authors wish to thank A, B, and C. This work was supported in part by
% a grant from XYZ.}

%\bibliographystyle{abbrv}

\bibliographystyle{abbrv-doi}

\bibliography{template}

\end{document}